\documentclass[10pt,aps,nofootinbib,
amsmath,amssymb,twocolumn,
preprintnumbers, 
showpacs,
raggedbottom, 
superscriptaddress,
prd,
floatfix]{revtex4-2}

\usepackage{cases}
\usepackage{pgf,tikz}
\usepackage{graphicx}

\usepackage{hyperref}
\usepackage[nameinlink]{cleveref}
\crefdefaultlabelformat{#2\textup{#1}#3}
\creflabelformat{equation}{#2\textup{#1}#3}

\newcommand{\ud}{\mathrm{d}}
\newcommand{\pd}{\partial}

\newcommand{\norm}[1]{\left|#1\right|}
\newcommand{\lrp}[1]{\left(#1\right)} 
\newcommand{\lrb}[1]{\left[#1\right]} 
\newcommand{\lra}[1]{\langle#1\rangle} 

\newcommand{\lam}{\lambda}

\newcommand{\al}{\alpha}
\newcommand{\be}{\beta}
\newcommand{\sig}{\sigma}

\newcommand{\diff}[2]{\frac{\ud #1}{\ud #2}}

\newcommand{\hf}{\frac{1}{2}}

\newcommand{\const}{\text{const}}

\newcommand{\MeV}{\text{MeV}}
\newcommand{\GeV}{\text{GeV}}

\newcommand{\hc}{\text{h.c.}}
\newcommand{\fmic}{\text{fm}^{-3}}
\newcommand{\gcm}{\text{g}/\text{cm}^3}
\newcommand{\dptil}[1]{\frac{\ud^3 #1}{2E_{#1}(2\pi)^3}}

\renewcommand{\vec}{\mathbf}

\newcommand{\eden}{\mathcal{E}}

\renewcommand{\vec}{\mathbf}

\begin{document}

\title{Dark Lepton Superfluid in Proto-Neutron Stars}

\newcommand{\INT}{Institute for Nuclear Theory, University of Washington, Seattle, WA 98195}
\newcommand{\UW}{Department of Physics, University of Washington, Seattle, WA 98195}

\author{Sanjay Reddy}
\email{sareddy@uw.edu}
\affiliation{\UW}
\affiliation{\INT}

\author{Dake Zhou}
\email{zdk@uw.edu}
\affiliation{\UW}
\affiliation{\INT}

\preprint{INT-PUB-21-016}

\date{\today}

\begin{abstract}
We find that sub-GeV neutrino portal bosons that carry lepton number can condense inside a proto-neutron star (newly born neutron star). These bosons are produced copiously and form a Bose-Einstein condensate for a range of as yet unconstrained coupling strengths to neutrinos. The condensate is a lepton number superfluid with transport properties that differ dramatically from those encountered in ordinary dense baryonic matter. We discuss how this phase could alter the evolution of proto-neutron stars and comment on the implications for neutrino signals and nucleosynthesis. 
\end{abstract}

\maketitle

\section{Introduction}

Hidden sectors (HS) that couple to the standard model (SM) often contain a rich spectrum of particles that can constitute dark matter (DM) \cite{Battaglieri:2017aum}. 
Their self-interactions and their interactions with the SM lead to interesting phenomenology and strategies for detection. 
In particular, scenarios where HS particles interact strongly with neutrinos has sparked recent interest \cite{Batell:2017cmf, Berryman:2018ogk, Brune:2018sab, Berlin:2018ztp, deGouvea:2019qaz, Kreisch:2019yzn, Blinov:2019gcj, Bally:2020yid}. These interactions can generate stronger self-interactions between neutrinos mediated by new degrees of freedom \cite{Kreisch:2019yzn,Blinov:2019gcj} and have been invoked to explain hints of anomalies in neutrino oscillation experiments \cite{Aguilar:2001ty, Aguilar-Arevalo:2018gpe} and puzzling cosmological observations  \cite{Bernal:2016gxb,Riess:2018kzi,Riess:2019cxk}.  
In this article we consider HS bosons that carry lepton number $L$ but is otherwise uncharged with respect to the SM gauge group and demonstrate its novel implications in extreme astrophysical environments.

A newly born neutron star, also called a proto-neutron star (PNS), provides a unique laboratory to study DM coupled to neutrinos. A PNS born in the aftermath of core-collapse supernovae explosion is hot ($T\simeq 30~\MeV$) and contains large densities ($\rho \simeq 5\times 10^{14}~\gcm$) of baryons and leptons. Detailed simulations have shown that neutrinos are trapped inside the PNS, and that their diffusion takes $\tau_\text{diff}\sim~$tens of seconds \cite{Woosley:1986ta, Burrows:2012ew}. During this time, the hot and dense PNS matter supports a large chemical potential for lepton number ($\mu_L=\mu_{\nu_e}\simeq 100-200~\MeV$) and a corresponding large excess of electron neutrinos. These conditions, which are not realized in any other environment in the universe, presents an opportunity to study DM candidates that carry lepton number.

\begin{figure}
\centering
\includegraphics[width=\linewidth]{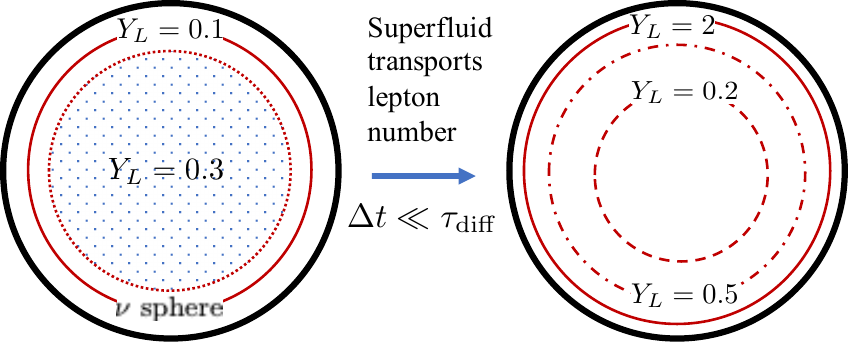}
\caption[Schematics for Lepton number transport by the superfluid]
{Superfluid transport of lepton number due to $\phi$ condensation rapidly quenches the spatial gradients in the lepton number chemical potential. It transforms the initial state with an approximately constant lepton fraction $Y_L\simeq 0.3$ (depicted on the left) to the state on the right.}
\label{fig:schematic}
\end{figure}

In this article we show that lepton number scalars, denoted by $\phi$, will have important implications for PNSs.
For a wide range of parameter space compatible with existing constraints, $\phi$ particles are rapidly produced in the PNS and thermalize to form a Bose-Einstein condensate. The condensate is a superfluid that transports lepton number from the core to the surface of PNS on timescales that are negligible compared to neutrino diffusion. This alters the radial distribution of lepton number (see ~\cref{fig:schematic}) and modifies the composition and neutrino transport properties of dense matter inside the PNS. 

We introduce the lepton number scalar $\phi$ through a minimal model defined by the low-energy effective Lagrangian \cite{Berryman:2018ogk}
\begin{equation}\label{eq:effL}
\mathcal{L}_\text{eff, int} \supset - \frac{g_{\al\be}}{2}\nu_\al\nu_\be\phi^* + \hc  -m_\phi^2\phi^*\phi- \frac{\lambda}{4}(\phi^*\phi)^2. 
\end{equation}
Here, $g_{\al\be}$ is the coupling between left-handed Weyl neutrinos $\nu_{\al,\be}$ and the scalar $\phi$ that carries lepton number 2. $\alpha,\beta$ are flavor indices and the quartic coupling $\lambda$ characterizes the strength of self-interaction among $\phi$'s. 
In general, $g_{\al\be}$ may involve all 3 flavors $e,\mu,\tau$. 
However, in what follows we focus on the flavor conserving coupling to electron neutrinos $g_{ee}=g$, and comment on the possible role of couplings to $\mu$ and $\tau$ neutrinos in our concluding remarks.

The coupling $g$ may arise after electroweak symmetry breaking from the effective operator $(l\tilde H)(l\tilde H)\phi^*/\Lambda^2$ \cite{Burgess_1994, Berryman:2018ogk}, where $l=(\nu_e,e)^T$ is the SM lepton doublet, $\tilde H=i\sig_2 H^*$ is the flipped SM Higgs doublet, and $\Lambda$ is the high-energy scale associated with new physics.
Since the Weinberg operator $(l\tilde H)(l\tilde H)$\cite{Weinberg:1979sa} carries lepton number 2, the Lagrangian \cref{eq:effL} conserves $L$. 
This effective operator is one of few dimension six operators that generates neutrino self-interactions.
In contrast to conventional majoron models where the coupling is highly suppressed by neutrino Majorana masses \cite{Chikashige:1980ui,Gelmini:1980re},  in this scenario, the natural strength for the coupling $g \simeq (v/\Lambda)^2$ ($v=246~\GeV$ is the electroweak scale) is expected to be large from the model building perspective \cite{Burgess_1994, Berryman:2018ogk}.

Laboratory constraints on the coupling $g_{\alpha\be}$ are flavor dependent. For $g\equiv g_{ee}$, limits from exotic meson decays demand $g\lesssim10^{-2}$ when $m_\phi\lesssim$ \GeV, and double beta-decay experiments require $ g\lesssim 10^{-4}$ when $m_\phi \lesssim \MeV$ \cite{Pasquini:2015fjv,Berryman:2018ogk, deGouvea:2019qaz}.
Astrophysical and cosmological bounds on the parameter space are also quite stringent. 
For $m_\phi \lesssim 10$~\MeV~and $g \gtrsim 10^{-10}$, a thermal population of $\phi's$ alters the expansion history of the early universe during and after weak decoupling. This spoils the concordance between the predictions of Big Bang nucleosynthesis (BBN) and observations of light element abundances \cite{Cyburt:2015mya,Blinov:2019gcj}, and  is also in tension with the observed cosmic microwave background (CMB) spectrum \cite{Kreisch:2019yzn}.
The supernovae cooling bound based on SN1987a -- the core-collapse supernova occurred in the Large Magellanic Cloud in 1987 -- excludes $g^{-11}\lesssim g \lesssim 10^{-6}$ when $m_\phi\lesssim 50$ MeV \cite{Heurtier:2016otg,Brune:2018sab}. This constraint arises because the duration of the observed neutrino signal is shortened when free streaming weakly interacting particles rapidly cool the PNS core subsequent to their thermal production \cite{Raffelt:1987yt}.

In this article, we will focus on the, as yet, unconstrained region for $\phi$'s with mass $m_\phi\gtrsim 10$ \MeV~ and a coupling to electron neutrinos in the range $10^{-6}\lesssim g \lesssim 10^{-2}$ to show that novel phenomena arise in the PNS that can probe new physics at  energy scale $\Lambda \sim 1-10^{4}$ TeV.\\

\section{$\phi$ condensation in the PNS}
As noted earlier, the PNS is characterized by high baryon number density $n_B \simeq 1-3~n_0$, where $n_0=0.16~\fmic$ is the nuclear saturation density, high temperature $T\sim 20-60$~ MeV, and a lepton fraction $Y_L=n_L/n_B \simeq 0.3$. In this hot and dense environment lepton number scalars are produced rapidly through $\nu\nu\rightarrow \phi$. The rate of production is given by
\begin{multline}\label{eq:prodrate}
\diff{n_{\nu\nu\rightarrow \phi}}{t} =\int\dptil{p_1}\dptil{p_2}\dptil{k} \\
\times \frac{\norm{\mathcal{M}_{\nu\nu\rightarrow \phi}}^2}{2}
 f_{1} f_{2} (2\pi)^4\delta^{4}(p_1+p_2-k)\,,
\end{multline}
where $\norm{\mathcal{M}_{\nu\nu\rightarrow \phi}}^2=4g^2(p_1\cdot p_2)$ is the square of the tree level matrix element, and $f_{{1,2}}=1/\lrp{1+\exp\lrb{(E_{p_{1,2}}-\mu_L)/T}}$ are the Fermi-Dirac distribution functions for neutrinos. For $g> 10^{-6}$ and $T\simeq 10$ MeV, the timescale for transferring lepton number to the $\phi$ particles $\tau_\text{prod}\simeq {n_L}/{\dot n_{\nu\nu\rightarrow \phi}} \ll\tau_\text{diff}$.
Further, most of the $\phi$'s produced have short mean free paths due to the inverse decay process hence are trapped and subsequently thermalize with the star.
Qualitatively, the relaxation timescale can be calculated as 
\begin{equation}\label{eq:relaxation}
\begin{aligned}
 \frac{1}{\tau}=\frac{1}{2E_k g_k}\int\dptil{p_1}\dptil{p_2}\\
\times \frac{\norm{\mathcal{M}_{\nu\nu\leftrightarrow \phi}}^2}{2}
 f_{1} f_{2} (2\pi)^4\delta^{4}(p_1+p_2-k),\\
\end{aligned}
\end{equation}
where $g_k=1/(\exp((E_k-\mu_\phi)/T)-1)$ denotes the Bose-Einstein distribution function of $\phi$'s. For a simple estimate we set all the distribution functions to $1$ and find that the mean free path of $\phi$'s
\begin{equation}\label{eq:trapping}
\begin{aligned}
\lambda_\phi = v \tau
\sim 10^{-8}~\lrp{\frac{10^{-3}}{g}}^2\lrp{\frac{50~\text{MeV}}{m_\phi}}^2 \lrp{\frac{E_\phi}{60~\text{MeV}}}
~\text{km}\,.
\end{aligned}
\end{equation}

Consequently, for the range of couplings considered in this study, $\phi$'s produced in PNS are in chemical and thermal equilibrium with neutrinos. 

Since $\phi$ carries two units of lepton number, its chemical potential $\mu_\phi=2 \mu_L$. When $\mu_\phi> m_\phi$, there will be a macroscopic occupation of $\phi$'s in the zero-momentum state, and the ground state is a Bose-Einstein condensate. In the PNS where $\mu_L \simeq 200$ MeV at early times, we can expect condensation for $m_\phi\lesssim 2\mu_{L,max}\simeq 400$ MeV.

In the absence of repulsive forces between $\phi$'s, the density of the condensate grows rapidly in the core of the PNS. When the total mass of bosons in the core exceeds $M_\text{crit} \simeq m^2_\text{pl}/m_\phi \approx (100~ \MeV/m_\phi)~10^{-18}~M_\odot$ the bosons collapse to form a black hole. The black hole once formed is expected to subsequently consume the entire star \cite{McDermott:2011jp,Jamison:2013yya}. Thus, the very existence of of neutron stars rules out such scenarios and viable models for $\phi$ must include repulsive self-interactions. Even a small quartic coupling $\lambda$ in Eq.~\ref{eq:effL} alters the effective potential for $\phi$'s \cite{Haber:1981ts,Benson:1991nj,Kapusta:2006pm,Bellac:2011kqa}
\begin{equation}
V_\text{eff}(\phi)=(m_\phi^2-\mu_\phi^2)\phi^*\phi+\frac{\lam}{4}(\phi^*\phi)^2\,,
\end{equation}
and provides the necessary stabilization. When condensation occurs the finite vacuum expectation value (vev) of the scalar field that minimizes the effective potential above is given by
\begin{equation}\label{eq:vev}
\lra{\phi}=\sqrt{2(\mu_\phi^2-m_\phi^2)/\lambda}\,.
\end{equation}
This vev breaks the global $U(1)_L$ symmetry and leads to superfluidity.

In the symmetry broken phase the scalar excitation spectrum is modified. The two degrees of freedom associated with the complex $\phi$ field manifest as fluctuations around the vev. Denoting the massless Goldstone mode as $J$, and the massive mode as $\sig$, in the condensed phase we can write $\phi=(f+\sig+iJ)/\sqrt{2}$, where $f= \sqrt{2}\lra{\phi}$. 
The dispersion relations for the goldstone mode ($\omega_-^\phi$) and the massive mode ($\omega_+^\phi$) are \cite{Haber:1981ts,Benson:1991nj,Kapusta:2006pm,Bellac:2011kqa}
\begin{equation}\label{eq:Bexcitations}
\omega_{\pm}^\phi=\sqrt{E_p^2+\mu_\phi^2+\frac{\lambda f^2}{2}\pm\sqrt{\frac{\lambda^2f^4}{16}+\mu_\phi^2(4E_p^2+2\lambda f^2)}}\,,
\end{equation}
where $E_p=\sqrt{p^2+m_\phi^2}$.

The propagation of neutrinos is also modified in the superfluid phase because $\nu$'s directly couple to the condensate. When $\phi$ acquires a vev, the Yukawa term in \cref{eq:effL} becomes
\begin{equation}\label{eq:YukawaL}
\mathcal{L}_\text{Yukawa}\supset \frac{g}{2\sqrt{2}}\lrp{f+\sigma+iJ}\nu\nu+\hc.
\end{equation}
The term proportional to $f$, which would have led to neutrino Majorana mass in the vacuum ($\mu_L=0$), leads to pairing of electron neutrinos near their Fermi surface. This BCS-like pairing introduces a gap $\Delta=g\lra{\phi}$ in the neutrino excitation spectrum 
\begin{equation}\label{eq:dispersion_nu}
\omega_{\pm}^\nu=\sqrt{\Delta^2+(p\pm\mu_L)^2}\,.
\end{equation}
For the viable range of coupling $g\lesssim10^{-2}$, this gap is negligible compared to the temperature in PNS.

The grand potential per unit volume for neutrinos is
\begin{multline}\label{eq:Omega_nu}
\Omega_\nu = \int\frac{\ud^3 p}{(2\pi)^3}\left[\hf(\omega_+^\nu+\omega_-^\nu) \right.\\
\left. + T\log\lrp{1+ e^{-\omega_+^\nu/T}}+T\log\lrp{1+ e^{-\omega_-^\nu/T}}\right],
\end{multline}
and for $\phi$'s is
\begin{multline}\label{eq:Omega_phi}
\Omega_\phi = \frac{f^2}{4}(m_\phi^2-\mu_\phi^2)- \int\frac{\ud^3 p}{(2\pi)^3}\left[\hf(\omega_+^\phi+\omega_-^\phi) \right.\\
\left. + T\log\lrp{1- e^{-\omega_+^\phi/T}}+T\log\lrp{1- e^{-\omega_-^\phi/T}}\right].
\end{multline}
The first term on the RHS of ~\cref{eq:Omega_phi} accounts for the condensate, and the second term contains the contribution from thermal excitations. 
UV divergences in \cref{eq:Omega_nu,eq:Omega_phi} are regulated by imposing a momentum cutoff while demanding the Goldstone mode remains gapless with the vev given in \cref{eq:vev}.

\begin{figure}[!htpb]
\centering
\includegraphics[width=\linewidth]{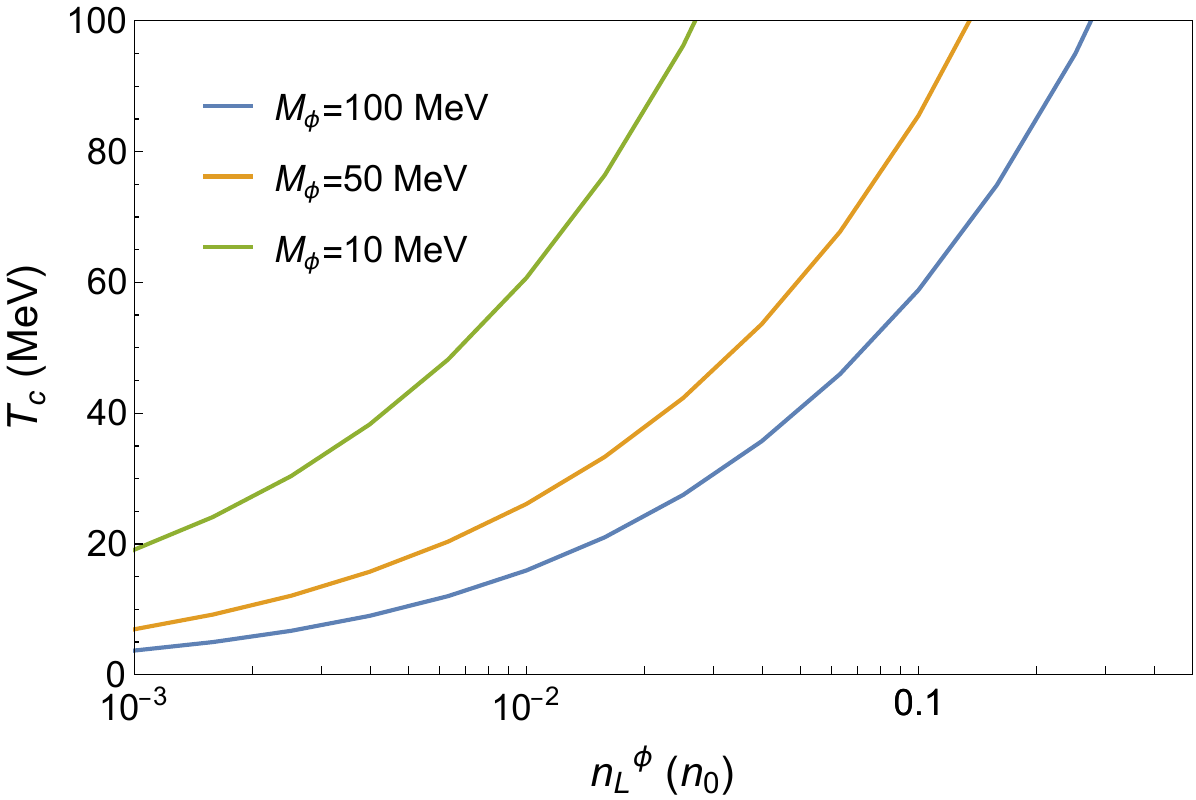}
\caption{Critical temperature of the superfluid in a gas comprised solely of $\phi$'s. }
\label{fig:Tcrit}
\end{figure}
In \cref{fig:Tcrit} we show the critical temperature as a function of the total lepton number density carried by $\phi$'s:  $n_L^\phi=\partial\Omega_\phi/\partial \mu_L$. For $T>T_c$, the lepton number resides in the thermal population and the condensate amplitude vanishes $f=0$. Note that the lepton number density in ~\cref{fig:Tcrit} is in units of $n_0$, and in the PNS where the lepton fraction $Y_L \simeq 0.3$ we expect 
$n_L \simeq 0.1-0.3~n_B$, and $n_B\simeq 1-3~n_0$. Under these conditions the large $T_c$ indicates that condensation will persist even at the highest temperatures encountered in the PNS when the lepton fraction is large.   

By accommodating a large fraction of the lepton number, the condensate alters the composition of hot and dense matter in the PNS. At given $n_B$, $T$, and $Y_L$,
the baryonic and leptonic components are related by the conditions of beta-equilibrium and charge neutrality:
\begin{align}
\mu_n -\mu_p  &= \mu_e- \mu_L\,,
\label{eq:equilibrium} \\
n_p&=n_{e}\,.
\label{eq:charge_neutral}
\end{align}
Above, subscripts $n$, $p$, $e$ denotes neutrons, protons, and electrons. Number densities $n_i$ include anti-particle contributions, e.g., $n_e=n_{e^-}-n_{e^+}$. In this work we treat electrons as non-interacting relativistic Fermi gas, and neglect muons for simplicity. 

To obtain the composition of hot and dense matter in equilibrium we need to specify a baryonic equation of state
$\eden (n_n,n_p,T)$, the nuclear energy density as a function of the neutron and proton number densities, and the temperature. For a given $\eden (n_n,n_p,T)$, the nucleon chemical potentials are given by $\mu_n=(\partial \eden(n_n,n_p)/\partial n_n)_{n_p,T}$, and  $\mu_p=(\partial \eden(n_n,n_p)/\partial n_p)_{n_n,T}$. In what follows we employ the Skyrme energy functional $\eden(n_n,n_p,T)$ discussed in Ref.~\cite{Steiner:2004fi}.

\begin{figure}[!htpb]
\centering
\includegraphics[width=\linewidth]{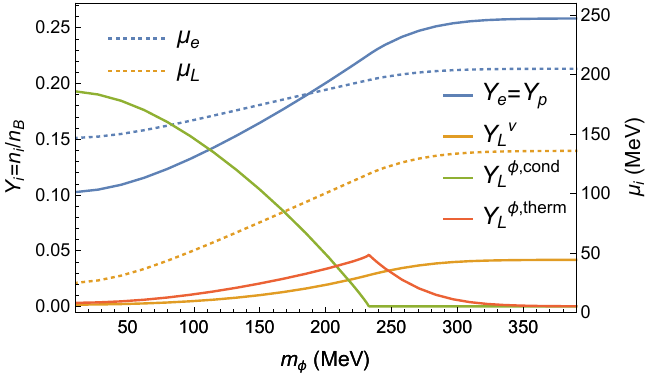}
\caption{PNS composition as a function of $m_\phi$ at $n_B=n_0$ assuming $T=20$ MeV and $Y_L=0.3$. The green curve and red curve represent lepton number carried by the condensate and thermal $\phi$'s respectively. We took $\lambda=1$. For $m_\phi\gtrsim 230$ MeV the $L$ budget can no longer support the superfluid in this scenario.}
\label{fig:Xs_Mphis}
\end{figure}

In ~\cref{fig:Xs_Mphis} we show the constituent particle fractions in the presence of $\phi$'s at $n_B=n_0$, $T=20$ MeV, and $Y_L=0.3$. The individual number densities are determined by chemical potentials $\mu_n$, $\mu_p$, $\mu_e$, and $\mu_L$ which are obtained by solving ~\cref{eq:equilibrium,eq:charge_neutral}, subject to the lepton number budget
$$n_L = n_BY_L= n_{e} + n_L^{\nu} + n_{L}^\phi,$$
where $n_L^{\nu}={\pd\Omega_\nu}/{\pd \mu_L}$ denote lepton number densities carried by $\nu$'s.

Effects of $\phi$'s on dense matter composition shown in ~\cref{fig:Xs_Mphis} can be understood by noting that the condensate amplitude decreases with increasing $m_\phi$.  
For a given value of $m_\phi$, the condensate amplitude is calculated using ~\cref{eq:vev} and we have set $\lambda=1$. 
For small $m_\phi$, the condensate amplitude is large and it accommodates a large mount of lepton number as shown by the green curve. Consequently, the populations of $e$'s (blue curve) and $\nu$'s (orange curve) are suppressed. Further, owing to charge neutrality \cref{eq:charge_neutral}, the condensate lowers the proton fraction and amplifies the isospin asymmetry of PNS matter at early times. 
At larger $m_\phi$, the lepton number in the condensate is smaller and vanishes when $m_\phi \ge 2\mu_L$. For the chosen parameters, this occurs at $m_\phi\simeq 230$ MeV. At finite temperature, a thermal bath of excitations also contributes to the lepton number density, and this contribution is shown by the red curve. It is interesting to note that for $m_\phi \lesssim 100$ MeV the $\phi$ condensation reduces the  proton fraction by about a factor 2.\\

\section{Implications for PNS Evolution}
\subsection{Lepton Number Transport} 
\begin{figure}[!htpb]
\centering
\includegraphics[width=\linewidth]{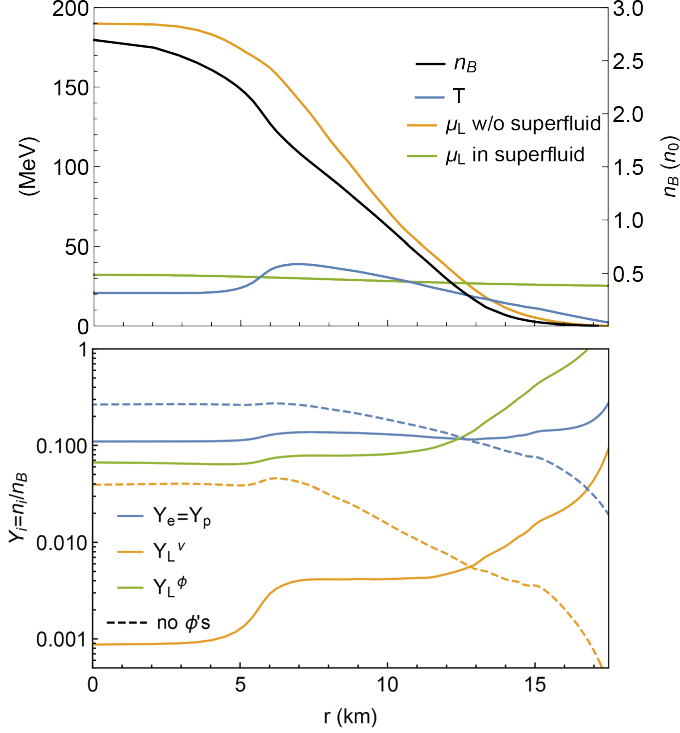}
\caption{Superfluid transports $L$ towards the surface of the star. Solid curves in the lower panel are obtained by distributing $N_{L,tot}\approx4\times10^{56}$ according to \cref{eq:metricMu} in a $1.4M_\odot$ star. The underlying  temperature, baryon number, and lepton chemical potential profiles are shown in the top panel.  We chose $m_\phi=50$ MeV, $g=10^{-3}$, and $\lam=1$.}
\label{fig:fix_mul}
\end{figure}

In the standard scenario, the evolution of the PNS and the associated neutrino signal is largely determined by the diffusion of neutrinos \cite{Burrows_1986,Pons:1998mm,Roberts:2016rsf}. Neutrino trapping during collapse leads to $Y_L \simeq 0.3$ in the bulk of the PNS, and $\mu_L\simeq 200$ MeV in the PNS core. The baryon density, temperature and lepton fraction profiles that roughly correspond to the state of the PNS at $t\simeq 1$ s (see Fig. 1 of Ref.~\cite{Roberts:2016rsf}) are shown in the top panel of ~\cref{fig:fix_mul}. The spatial gradient of $\mu_L$ drives deleptonization by neutrino diffusion and simulations suggest that it takes about $20$ seconds for $\mu_L$ to decrease substantially \cite{Roberts:2016rsf}. During this time, the lepton number current carried by neutrinos leads to heating in the core akin to Joule heating in ordinary conductors \cite{Burrows_1986}. As we discuss below, $\phi$ condensation fundamentally alters lepton number transport in PNS.   

A defining characteristic of a superfluid is that it supports dissipationless supercurrents that eliminate gradients in the chemical potential. Superflow redistributes the lepton number in the PNS on a short timescale $ \tau \ll \tau_{\rm diff}$ without the associated Joule heating. Consequently, the PNS is characterized by a constant lepton number chemical potential in the star's gravitational field
\begin{equation}\label{eq:metricMu}
\mu_L(r)\sqrt{g_{tt}(r)}=\mu_L^0=\const,
\end{equation}
where $\mu_L(r)$ is the local chemical potential, 
and $g_{tt}(r)$ is the time-time component of the metric tensor.
The value of $\mu_L^0$ is set by the total lepton number ($N_{L,tot}$) in the star which at early times is $N_{L,tot}\approx 3-5 \times10^{56}$.  

The bottom panel of ~\cref{fig:fix_mul} shows how the superfluid transport of lepton number alters the composition of the PNS. In the presence of the condensate, it ensures that the chemical potential $\mu_L(r)$ satisfies ~\cref{eq:metricMu}. To obtain metric functions we solved the Tolman–Oppenheimer–Volkoff equation assuming the Skyrme equation of state mentioned earlier. At fixed $N_{L,tot}$, curves shown in the top panel reveals that $\mu_L(r)$  in the core is greatly reduced relative to the standard scenario, and is enhanced in the surface regions.  Thus, in stark contrast to the standard scenario, the electron and the neutrino fractions are reduced in the core, and increase towards the surface of the PNS. A comparison between the solid and dashed curves in the lower panel reveals that the electron (and proton) fraction is reduced by about a factor of 2 in the core, and is enhanced in the low-density surface regions. The rapid increase in the neutrino fraction at intermediate radius is due to thermal effects and is sensitive to the  chosen temperature profile. We experimented with a few different initial profiles of $T$ and $\mu_L$, and found that qualitative features of lepton number transport by the superfluid seen in \cref{fig:fix_mul} are robust. The results shown in ~\cref{fig:fix_mul} are obtained by setting $m_\phi=50$ MeV and $\lam=1$. We find that the superfluid can extend to large radii at early times for $m_\phi \lesssim 50$ MeV. 

\subsection{Energy Transport}
While condensation greatly accelerates deleptonization in the inner PNS core by superfluid transport of lepton number, in what follows we shall argue that it has modest effect on energy transport. The temperature gradient in the PNS drives neutrino diffusion, and all flavors of neutrinos contribute to the energy flux. We can expect the transport of energy in the condensate to differ from the standard scenario because the $\nu_e$ and $\bar{\nu}_e$ mean free paths are altered. In particular, we find that the mean free path due to the Cherenkov process $\nu\rightarrow\nu J$
\begin{multline}\label{eq:cherenkovrate}
\frac{1}{\lambda(E_p)} =\frac{1}{E_p}\int\dptil{p'}\dptil{k'}\\
\times \norm{\mathcal{M}_{\nu\rightarrow \nu J}}^2
 (1-f_{p'}) (2\pi)^4\delta^{4}(p-p'-k')\,,
\end{multline}
is significantly shorter than other processes previously considered. In ~\cref{eq:cherenkovrate}, $p=(E_p,\vec{p})$ is the four-momentum of the initial neutrino, and $p'$, $k'$ are the four-momenta of the neutrino in the final state and the Goldstone boson, respectively. This process is kinematically feasible because the Goldstone mode (c.f. \cref{eq:Bexcitations}) exhibits a linear dispersion with velocity $v_J=\sqrt{(\mu_\phi^2-m_\phi^2)/(3\mu_\phi^2-m_\phi^2)}<1$ at small momenta. However, with increasing momenta, when $|\vec{k'}| \gtrsim \lra{\phi}$ the dispersion relation resembles the vacuum mode of $\phi$'s and the process is kinematically forbidden. We therefore require that $|\vec{k'}|\lesssim\lra{\phi}$ in ~\cref{eq:cherenkovrate}.

\begin{figure}[!htpb]
\centering
\includegraphics[width=\linewidth]{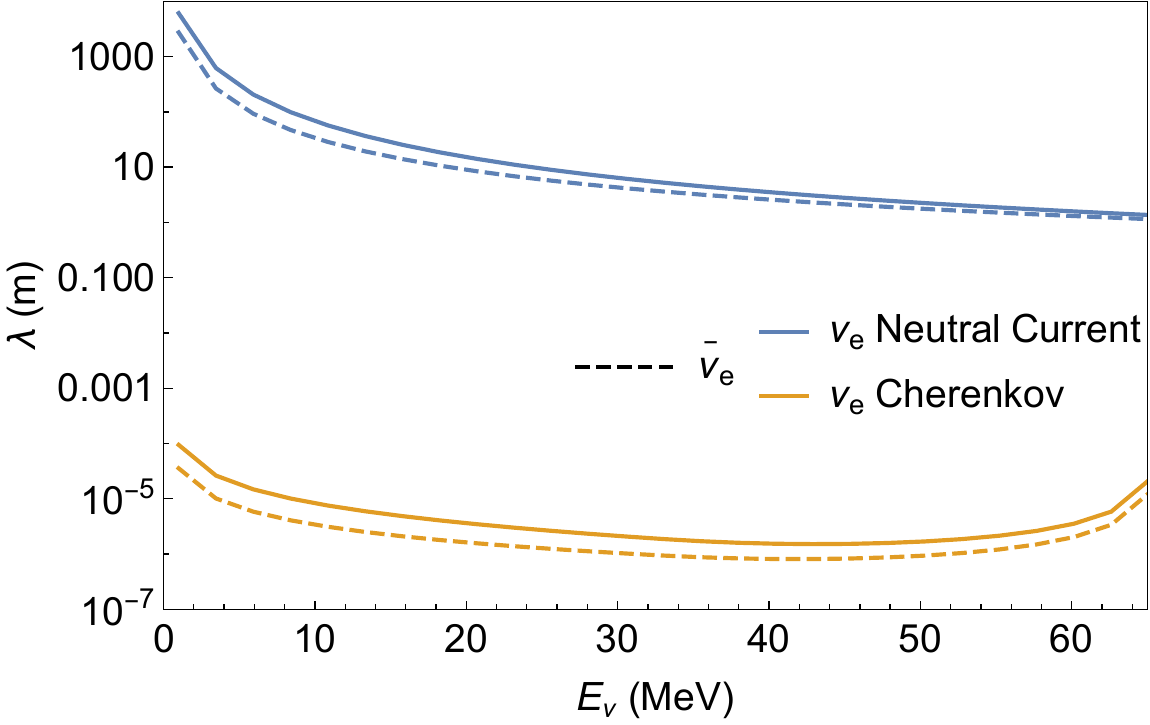}
\caption{Neutrino mean free paths at $n_B=n_0$ in the condensate depicted in \cref{fig:fix_mul}, where $T\approx27$ MeV and $\mu_L\approx30$ MeV. As $E_\nu$ increases, $\nu\rightarrow\nu+J$ becomes forbidden at sufficiently large Goldstone momenta (see main text), resulting in the rising tail for the Cherenkov mean free path.} 
\label{fig:MFP}
\end{figure}
\Cref{fig:MFP} shows the Cherenkov mean free paths for neutrinos and anti-neutrinos as functions of their energy. For reference, the mean free path due to neutral current reactions  are also shown. These mean free paths are calculated assuming non-relativistic nucleons under the influence of mean field potentials \cite{Reddy:1997yr} obtained from the Skyrme model discussed earlier. The drastic shortening, by orders of magnitude, of the $\nu_e$ and $\bar{\nu}_e$ mean free paths suggests that they will not contribute to energy transport in the PNS. However, in our scenario since $\mu$ and $\tau$ neutrinos do not couple to the Goldstone modes, they will dominate energy transport. Thus, we expect condensation to reduce the energy flux in the PNS by $1/3$. This would increase the corresponding cooling timescale by about 50\%.

\subsection{Evolution of the Condensate}
The radius at which the superfluid terminates, which we denote as $R_L$, is sensitive to the net lepton number in the star and $m_\phi$. From ~\cref{fig:fix_mul}, we see for $m_\phi \simeq 50$ MeV the condensate can extend to large radii at early times. As the PNS evolves and looses lepton number, the superfluid boundary will recede. 
Since the neutrino mean free path is generally very large at the low temperatures encountered at large radii, the edge of the superfluid is expected to retract to a location inside the conventional neutrino sphere quickly.
Here we shall show that the timescale associated with subsequent evolution is on the order of 10s, and is set by the cooling timescale of the PNS. 

The electron neutrino mean free path in the region $r>R_L$ due to the reaction $\nu_e \nu_e \rightarrow \phi$ given by
\begin{multline}
\lambda_{\nu_e}^{-1}(E_1) =\frac{1}{2E_1}\int\dptil{p_2} f_2~ \dptil{k} \\
\times \norm{\mathcal{M}_{\nu\nu\rightarrow \phi}}^2
 (2\pi)^4\delta^{4}(p_1+p_2-k)\,,\\
 =\frac{g^2m_\phi^2T}{8\pi E_1^2}\log\lrb{1+\eta_{\nu_e} \exp\lrp{-\frac{m_\phi^2}{4E_1 T}}}
 \label{eq:lambda_recession}
 \end{multline} 
is very short at early times due to the high temperature encountered in the PNS. Here, $\eta_{\nu_e}=\exp{(\mu_L/T)} \simeq 1$ is the neutrino degeneracy factor, and $E_1$ is the energy of the neutrino.

\begin{figure}[!htbp]
\centering
\includegraphics[width=0.95\linewidth]{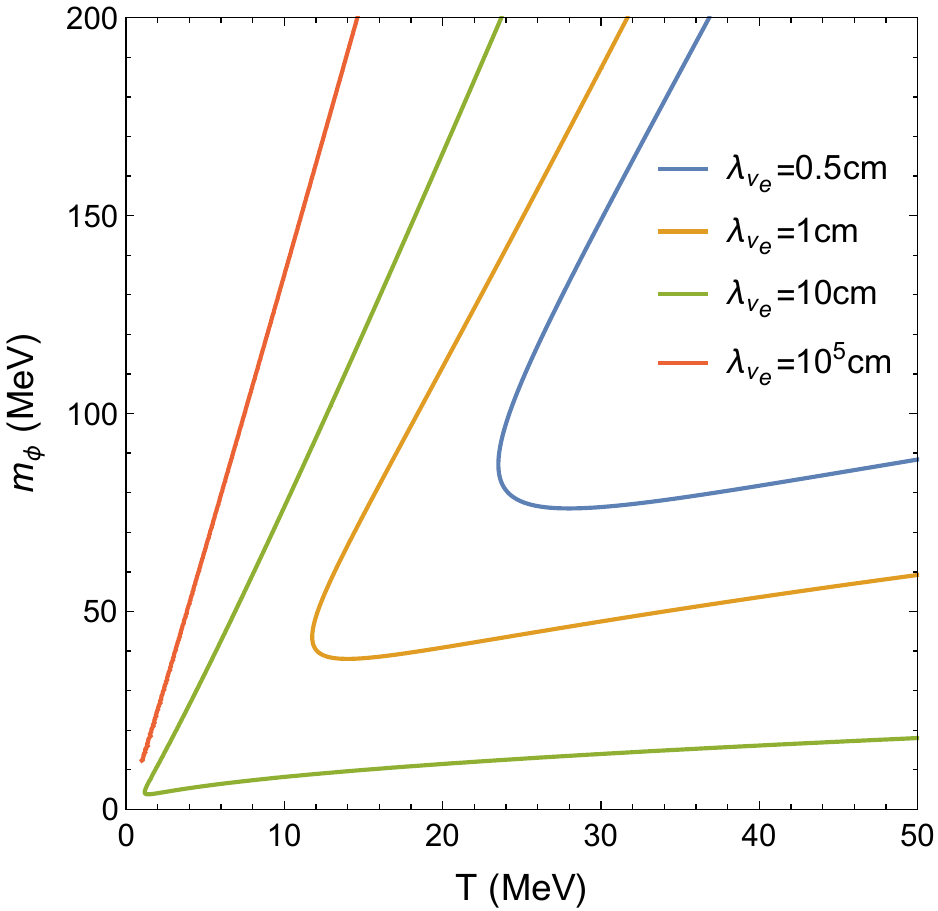}
\caption{Contours of constant neutrino mean free path. Results shown were obtained for $g=10^{-5}$, $\eta_{\nu_e}=1$ and neutrino energy $E_1=3T$.}
\label{fig:neutrino_mean_free_path}
\end{figure}
Fig.~\ref{fig:neutrino_mean_free_path} shows contours of constant neutrino mean free path for $g=10^{-5}$ and neutrino energy $E_\nu=3T$ in the $g-m_\phi$ plane. At the high temperatures realized in the PNS at early times, the $\nu_e$ mean free path is significantly shorter than the mean free paths of $\mu$ and $\tau$ neutrinos that contribute to cooling.  However, the $\nu_e$ mean free path increases exponentially at low temperature. The short $\nu_e$ mean free paths at high temperature and its steep temperature dependence at low temperature ($T/m_\phi\lesssim 0.1$) implies that lepton number transport in the region $r>R_L$ will accelerate rapidly as the PNS cools to $T\lesssim m_\phi/10$. Consequently, we can expect the timescale associated with the recession of the superfluid boundary to be set by the PNS cooling timescale.  Earlier studies find that PNS cooling occurs on a timescale of order 10s, and since lepton superfluidity has only modest effect on energy transport we can expect the superfluid to persist during this timescale. However, given the dynamical nature of this process, feedback can be important for neutrino transport and self-consistent PNS simulations \cite{Pons:1998mm, Burrows:2012ew} will be needed to draw definitive conclusions about the extent of lepton superfluidity in the PNS and its evolution with time.  

\subsection{Neutrino Decoupling} 
For small $m_\phi$ and large coupling $g$, the reactions $\nu_e \nu_e \leftrightarrow \phi$ and $\bar{\nu}_e \bar{\nu}_e \leftrightarrow \phi^*$ in the vicinity of the neutrino-sphere can alter the energy spectrum of electron neutrinos emitted from the PNS. Here, we will make a few simplifying assumptions to identify regions of the $m_\phi-g$ parameter space that will impact neutrino decoupling. First, we shall assume that all neutrinos decouple from a neutrino-sphere of radius $R_\nu$. Our second assumption is that the propagation of neutrinos in the vicinity of the neutrino-sphere is dominated by their interaction with other neutrinos. This will depend sensitively on the degree of anisotropy of the neutrino radiation field in this region. To model the anisotropy we employ the neutrino light-bulb model \cite{Duan:2010bg}. This model is based on simple geometric considerations, and predicts that the distribution of neutrinos at a distance $r> R_\nu$ is given by   
\begin{equation}
f(E_\nu,r,\mu) = f_{\nu}(E_\nu) \xi^2 \frac{\Theta(\mu-\sqrt{1-\xi^2})}{1-\sqrt{1-\xi^2}}\,,
\label{eq:lightbulb}
\end{equation} 
where $\xi=R_\nu/r<1$, $\mu=\cos{\theta}$ and $\theta$ is the angle measured with respect to the radial direction.  Here, $f_\nu(E_\nu)$ is the Fermi-Dirac distribution function at the neutrino-sphere which is characterized by temperature $T_\nu$. The Heaviside step function ensures that the neutrino flux becomes increasingly forward peaked for $r> R_\nu$, and the factor $\xi^2$ accounts for the dilution.

The inverse mean free path of electron neutrinos with energy $E_1$ propagating in an anisotropic neutrino distribution can be calculated similar to \cref{eq:lambda_recession}, with $f_2$ replaced by $f(E_\nu,r,\mu)$ from \cref{eq:lightbulb}. It is found that $\lambda_{\nu_e}(E_1,r)= \lambda_{\nu_e} (E_1)/G(\xi)$ where 
\begin{equation}
\lambda^{-1}_{\nu_e,\bar{\nu}_e}(E_1)=
\frac{g^2 m_\phi^2 T}{8\pi E^2_1}~ \exp{\left(-\frac{m_\phi^2}{4E_1T_\nu}\right)}\,,
\label{eqn:lambda_nubar_decouple}
\end{equation}
and
\begin{multline}
G(\xi)=\frac{\xi^2}{1-\sqrt{1-\xi^2}} \exp{\left[-\frac{m_\phi^2}{4E_1T_\nu} \left(\frac{1+\sqrt{1-\xi^2}}{1-\sqrt{1-\xi^2}}\right) \right]}\,. 
\end{multline}

In obtaining the results above we have approximated the Fermi-Dirac distribution function with that of Boltzmann.

This is justified since PNS simulations indicate that the neutrino chemical potential is small in the outer low-density regions of the PNS and $\eta_{\nu_e}=e^{\mu_L/T}\approx 1$ in the vicinity of the neutrino-sphere in the standard scenario. 
The presence of a lepton superfluid that extends to regions close to the neutrino-sphere could enhance $\eta_{\nu_e}$ at early times. However, the discussion in the preceding sub-section suggests that the superfluid would recede to regions of higher temperatures rapidly.
In what follows, we shall assume that  $\eta_{\nu_e}\simeq 1$ in the decoupling region.  
This suggests the $\nu_e$ and $\bar{\nu}_e$ mean free paths will be similar in the decoupling region.

\begin{figure}[!htbp]
\centering
\includegraphics[width=0.95\linewidth]{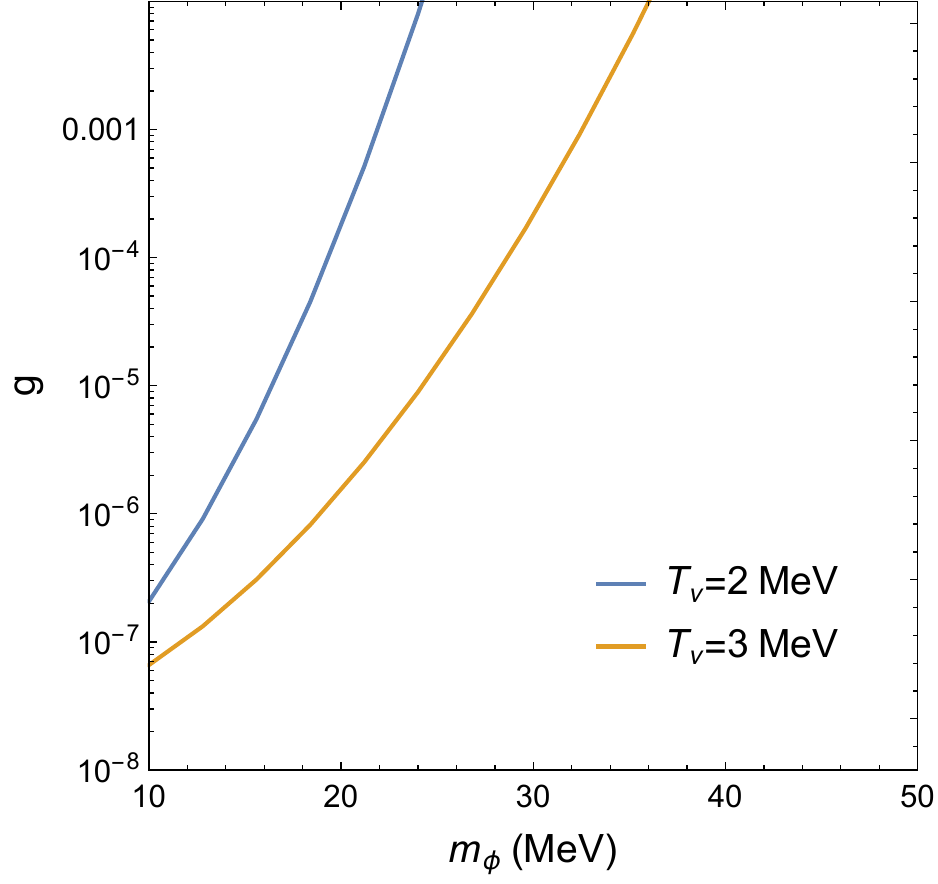}
\caption{Contours of constant decoupling temperature for electron anti-neutrinos with $L_{\bar{\nu}_e}=3 \times 10^{51}$ erg/s. In the disfavored region where $T\lesssim3$ MeV, $R_\nu\gtrsim15$ km is consistent with our assumption that the (anti-)neutrino decoupling is determined by (${\bar\nu}_e{\bar\nu}_e\leftrightarrow\phi^*$) $\nu_e\nu_e\leftrightarrow\phi$. }
\label{fig:decouple}
\end{figure}

To model the decoupling of $\nu_e$ and $\bar\nu_e$ we rely on the assumption that the emergent spectrum can be approximated by that of a blackbody. This will allow us to relate the neutrino decoupling radius and its temperature at a given luminosity. The decoupling radius $R_\nu \simeq [L_\nu/(\frac{7}{8}\hf4\pi \sigma_{\rm SB}T^4_\nu)]^{1/2}$, where $L_\nu$ is the neutrino luminosity,  $T_\nu$ is the decoupling temperature and $\sigma_{\rm SB}=1.02\times  10^{36}$ erg/MeV$^4$/cm$^2$/s is the Stefan-Boltzmann constant. The optical depth of a typical electron anti-neutrinos with thermal energy $E_\nu= 3 T_\nu$ is given by 
\begin{multline}
\label{eqn:nudecouple}
\tau_{\rm D}=\int^\infty_{R_\nu} \frac{\ud r}{ \lambda_{\bar{\nu}_e}(3T_\nu,r)}
=\frac{R_\nu}{\lambda_{\bar{\nu}_e}(3T_\nu)}~\int^1_0 d\xi\frac{G(\xi)}{\xi^2}\\
=\frac{R_\nu}{\lambda_{\bar{\nu}_e}(3T_\nu)}\int^1_0 \frac{\ud\xi}{1-\sqrt{1-\xi^2}} \\
\times\exp{\left[-\frac{m_\phi^2}{12T_\nu^2} \left(\frac{1+\sqrt{1-\xi^2}}{1-\sqrt{1-\xi^2}}\right) \right]} \,.
\end{multline}
Assuming that neutrino decoupling occurs at an optical depth $\tau_D=2/3$, and for a given luminosity, we can use ~\cref{eqn:nudecouple} to estimate the $\bar{\nu}_e$ decoupling temperature, $T_\nu$. In ~\cref{fig:decouple} we plot contours of constant $T_\nu$ in the $g-m_\phi$ plane. The results suggest that the reaction $\nu\nu\rightarrow \phi$ can influence the neutrino decoupling when  $m_\phi \lesssim 50$ MeV and $g \gtrsim 10^{-6}$. Further, since the analysis of neutrino events from SN1987a favors $T_\nu \gtrsim 3$ MeV~\cite{Loredo:2001rx}, the parameter space to the left of the contour labelled $T_\nu=3$ MeV is disfavored. We note that this analysis of neutrino decoupling applies to a wide range of models in which neutrinos couple to new scalars.

The results presented in this section imply that
when neutrino decoupling is dominated by the reactions $\nu_e \nu_e \leftrightarrow \phi$ and $\bar{\nu}_e \bar{\nu}_e \leftrightarrow \phi^*$, we should expect the $\nu_e$ and $
\bar{\nu}_e$ spectra to be similar. This could have important implications for supernova  nucleosynthesis because differences between $\nu_e$ and $
\bar{\nu}_e$ spectra are known to alter the proton fraction of the matter ejected from PNSs~\cite{Qian:1996xt}. If $\nu_e$ and $
\bar{\nu}_e$ have identical spectra, the composition of matter ejected will be proton-rich and would preclude r-process nucleosynthesis~\cite{Thielemann:2011}. Instead, these proton-rich conditions are expected to produce lighter heavy elements with $A<130$, such as Sr, Y, Zr~\cite{Qian:2008qs}. 

From ~\cref{eqn:lambda_nubar_decouple,eqn:lambda_nubar_decouple} we can also deduce that neutrino-neutrino interactions are exponentially sensitive to the neutrino energy when $E_1 \ll  m_\phi^2/4T_\nu$. This will likely distort the shape of the emerging neutrino spectrum, and enhance the flux of low energy neutrinos. This strong energy dependence also suggests that a more sophisticated treatment of neutrino transport is needed to derive robust constraints on $g$ and $m_\phi$ using SN1987a data. Such studies, which are beyond the scope of this article, would rely on solving the neutrino Boltzmann equation in the PNS which would also include self-consistently neutrino interactions with baryons and electrons in the decoupling region.

\section{Conclusion}\label{sec:concl}

Our main conclusion is that if lepton number scalars that couple to neutrinos exist in nature they will condense to form a superfluid in PNS. For the conditions realized inside PNS, condensation is favored for an interesting range of masses and couplings, and is likely to persist for several seconds during which lepton number remains trapped. We find that condensation dramatically alters the composition and transport properties of hot and dense PNS matter.

This first study of lepton number superfluidity in hot and dense matter suggests that the thermodynamic and transport properties are dramatically altered. The large spatial extent of the condensate, and shortened neutrino mean free paths in the condensed phase suggest that the energy spectrum and the luminosity of neutrinos could be affected, especially at early times.
However, identifying unique observable signatures in the neutrino signal will require self-consistent PNS simulations that include lepton number scalars. Such studies could address how neutrino data from SN1987a, and future galactic supernovae can help either discover or constrain a dark lepton sector. 
We also note that the condensation of lepton number scalars could occur in binary neutron star mergers as well where similar astrophysical conditions are realized.

Our analysis in this article neglected the diagonal and off-diagonal couplings of the scalar $\phi$ to $\mu$ and $\tau$ neutrinos. Their inclusions could be interesting as it provides additional degrees to accommodate lepton number. It could have an impact during the infall phase of core-collapse~\cite{Fuller:1988ega} and the PNS evolution studied here. These issues are being investigated and will be reported in future works.

Finally, we note that dark superfluid of other forms could have interesting implications in different astrophysical settings. For instance, an additional dark superfluid component is used to revive Modified Newtonian Dynamics in explaining the dark matter in galaxies and clusters of galaxies ~\cite{Berezhiani:2015bqa,Berezhiani:2015pia}. Recently, it is also hypothesized that a dark chiral condensate in the center of galaxies may resolve the ``core-cusp'' problem~\cite{Alexander:2019qsh,Alexander:2020wpm}.

\section*{Acknowledgment}
We thank Andre de Gouvea and George Fuller for useful comments on the manuscript, and Luke Roberts for helpful discussions. This work is supported by Grant No. DE-FG02-00ER41132 from the Department of Energy , and the Grant No. PHY-1430152 (JINA Center for the Evolution of the Elements), and PHY-1630782 (Network for Neutrinos, Nuclear Astrophysics, and Symmetries (N3AS)) from the National Science Foundation.

\bibliographystyle{apsrev4-2}

\appendix
\section{$\phi$ production rate}

Here we present an estimate for the $\phi$ production rate. Define dimensionless quantities $x_i=p_i/T$, $y=\mu_L/T$ and $z=m_\phi/T$, \cref{eq:prodrate} can be evaluated as
\begin{subequations}
\begin{equation*}
\dot{n}_\phi = \frac{g^2 m_\phi^2 T^2}{32\pi^3}
\int^\infty_0\ud x_1\int^\infty_{z/(4x_1)}\frac{\ud x_2}{[e^{x_1-y}+1][e^{x_2-y}+1]}
\end{equation*}\vspace{-5ex}
\begin{multline*}
\approx 10^{61}\ \text{s}^{-1}\text{km}^{-3}
\times \lrp{\frac{g}{10^{-3}}}^2\lrp{\frac{m_\phi}{50\text{ MeV}}}^2 \lrp{\frac{T}{30\text{ MeV}}}^2\\
\times\mathcal{I}(\mu_L, m_\phi,T).
\end{multline*}
\end{subequations}

\begin{figure}[!htbp]
\centering
\includegraphics[width=\linewidth]{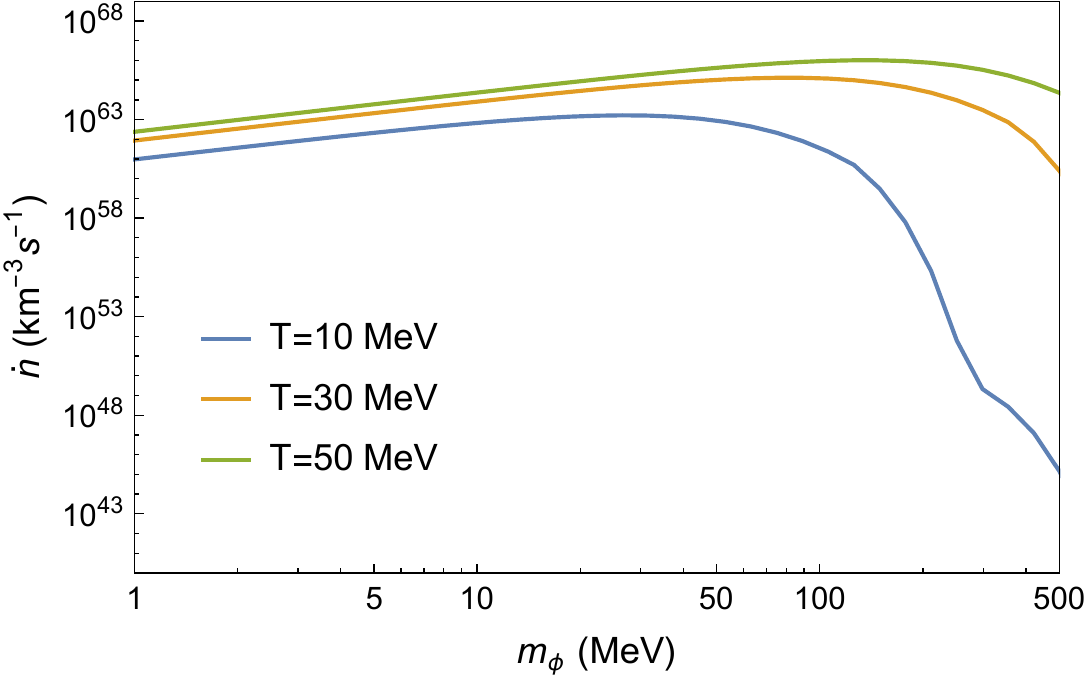}
\caption{The $\phi$ production rate assuming $\mu_L=0$ (a conservative bound).}
\label{fig:prod}
\end{figure}

Note that the production rate vanishes in the limit $m_\phi\rightarrow0$, since we ignored the neutrino masses. Above, $\mathcal{I}(\mu, m_\phi,T)$ denotes the result of the double integral in the first line. To obtain an conservative estimate we may treat neutrinos as non-degenerate. The result is shown in \cref{fig:prod}.

\end{document}